\documentclass[11pt,twoside]{article}
\input{epsf.sty}
\usepackage{mathrsfs}
\usepackage{amsmath}
\usepackage{amssymb}
\usepackage{graphicx}
\usepackage{subfigure}
\newtheorem{proposition}{Proposition}

\newtheorem{corollary}{Corollary}

\title{An integrable (2+1)-dimensional Camassa-Holm hierarchy with peakon solutions}
\author{Baoqiang Xia$^{1}$\footnote{E-mail address:
xiabaoqiang@126.com}, ~~Zhijun Qiao$^2$\footnote{E-mail address:
qiao@utpa.edu},
\\
$^{1}$School of Mathematics and Statistics, Jiangsu Normal
University,\\
Xuzhou, Jiangsu 221116, P. R. China\\
$^2$Department of Mathematics, University of Texas-Pan American, \\Edinburg, Texas 78541, USA
\\
}

\date{}
\begin{document}
\maketitle
\begin{abstract}
In this letter, we propose a  (2+1)-dimensional generalized Camassa-Holm (2dgCH) hierarchy with both quadratic and cubic nonlinearity.
The Lax representation and peakon solutions for the 2dgCH system are derived.

\vspace{0.5cm}


\end{abstract}
\newpage

\section{ Introduction}
In recent years, the Camassa-Holm (CH) equation \cite{CH}
\begin{eqnarray}
m_t-\alpha u_x+2m u_x+m_xu=0, \quad m=u-u_{xx},
\label{bCH}
\end{eqnarray}
has attracted much attention in the theory of integrable systems and solitons.
Since the work of Camassa and Holm \cite{CH}, various studies on this equation have remarkably been developed \cite{OR}-\cite{JR}. The most remarkable feature of the CH equation (\ref{bCH}) is that it admits peaked
soliton (peakon) solutions in the case of $\alpha=0$ \cite{CH,CH2}. A peakon is a weak
solution in some Sobolev space with corner at its crest.
The stability and interaction of peakons were discussed in several references \cite{CS1}-\cite{JR}.

In addition to the CH equation being an integrable model with peakon solutions,
other integrable peakon models have been found, including the Degasperis-Procesi (DP) equation \cite{DP1} whose
 Lax pair, bi-Hamiltonian formulation and peakon solutions were discovered in \cite{DP2,DP3},
the cubic nonlinear peakon equations \cite{Q1,NV1,HW1},
and a generalized CH equation (gCH) with both quadratic and cubic nonlinearity \cite{QXL}
\begin{eqnarray}
 m_t=\frac{1}{2}k_1\left[ m(u^2-u^2_x)\right]_x+\frac{1}{2}k_2(2 m u_x+ m_xu), \quad  m=u-u_{xx},\label{gCH}
\end{eqnarray}
where $k_1$ and $k_2$ are two arbitrary constants.
Through some appropriate rescaling, equation (\ref{gCH}) could be transformed to the one in the papers of Fokas and Fuchssteiner \cite{Fo,Fu}, where it was derived from  the  motion of a two-dimensional, inviscid,
incompressible fluid over a flat bottom.
In \cite{QXL}, the Lax pair, bi-Hamiltonian structure, peakons, weak kinks, kink-peakon interactional and smooth soliton solutions of equation (\ref{gCH}) are presented.

It is an interesting task to study the (2+1)-dimensional generalizations of the peakon equations. For example, in \cite{EP,EP2} the authors provided
a (2+1)-dimensional extension of the CH hierarchy, and they further studied the hodograph transformations and peakon solutions for their (2+1)-dimensional CH equation. In the present letter, we generalize the gCH equation (\ref{gCH}) to the whole integrable hierarchies in (1+1) and (2+1)-dimensions. We show that the gCH hierarchies admit Lax representations and construct a relation between the gCH hierarchies in (1+1) and (2+1)-dimensions. Moreover, we derive the single-peakon solution and the multi-peakon dynamic system for the (2+1)-dimensional gCH equation.

The letter is organized as follows. In section 2, we review the CH hierarchies in (1+1) and (2+1)-dimensions. In section 3, we present the gCH hierarchies in (1+1) and (2+1)-dimensions. In particular, we give their Lax representations. In section 4, we derive the peakon solutions to the $(2+1)$-dimensional gCH equation. Some conclusions and discussions are drawn in section 5.

\section{Overviews}

In this section, we review the (1+1) and (2+1)-dimensional CH hierarchies presented in \cite{Q3,EP,EP2}.
The new results we find are a relation between the CH hierarchies in (1+1) and (2+1)-dimensions and an isospectral Lax representations for the CH hierarchies.

\subsection{The CH hierarchies in (1+1) and (2+1)-dimensions}
Let us consider the Lenard operators pair \cite{CH}
\begin{eqnarray}
J=\partial_x m+m\partial_x, \quad K=\frac{1}{2}(\partial^3_x-\partial_x).
\end{eqnarray}
The Lenard gradients $b_{-k}$ are defined recursively by
\begin{eqnarray}
Kb_{-k}=Jb_{-k+1}, \quad Kb_0=0, \quad k\in\mathbb{Z^{+}}.
\end{eqnarray}
Taking an initial value $b_0=-\frac{1}{2}$, one may generate the negative CH hierarchy \cite{Q3}
\begin{eqnarray}
\left\{\begin{array}{l}
m_{t_{-n}}=Jb_{-n},\\
Kb_{-j}=Jb_{-j+1},
\end{array}\right.  \quad 1\leq j\leq n.
\label{CHH}
\end{eqnarray}
For $n=1$, (\ref{CHH}) becomes
\begin{eqnarray}
\left\{\begin{array}{l}
m_{t_{-1}}=(mb_{-1})_x+mb_{-1,x},\\
\frac{1}{2}(b_{-1,xxx}-b_{-1,x})=-\frac{1}{2}m_{x},
\end{array}\right.
\label{CHH1}
\end{eqnarray}
which is nothing but the CH equation (\ref{bCH}) with $\alpha=0$ \cite{CH}.
For $n=2$, we arrive at
\begin{eqnarray}
\left\{\begin{array}{l}
m_{t_{-2}}=(mb_{-2})_x+mb_{-2,x},\\
\frac{1}{2}(b_{-2,xxx}-b_{-2,x})=(mb_{-1})_x+mb_{-1,x},
\\
\frac{1}{2}(b_{-1,xxx}-b_{-1,x})=-\frac{1}{2}m_{x}.
\end{array}\right.
\label{CHH2}
\end{eqnarray}
In what follows, we call equation (\ref{CHH2}) the $2$-nd CH equation.
For the general $n$, we refer to (\ref{CHH}) as the $n$-th CH equation.

In \cite{EP,EP2}, the authors proposed a (2+1)-dimensional CH equation
\begin{eqnarray}
\left\{\begin{array}{l}
m_{t}=(mb_{-2})_x+mb_{-2,x},\\
\frac{1}{2}(b_{-2,xxx}-b_{-2,x})=m_y.
\end{array}\right.
\label{CH3d}
\end{eqnarray}
In general, a $(2+1)$-dimensional generalization of the CH hierarchy could be written as \cite{EP,EP2}
\begin{eqnarray}
\left\{\begin{array}{l}
m_{t_{-n}}=Jb_{-n},\\
Kb_{-j}=Jb_{-j+1},
\\
Kb_{-2}=m_y,
\end{array}\right.  \quad 3\leq j\leq n.
\label{CHH3d}
\end{eqnarray}
In \cite{EP,EP2}, the authors also studied the hodograph transformations and the peakon solutions of the (2+1)-dimensional CH equation.

\subsection{Lax representation}

Let
\begin{eqnarray}
U=\left( \begin{array}{cc} 0 & 1\\
 \frac{1}{4}+\lambda m &  0 \\ \end{array} \right),
 \quad
 V^{(-n)}=-\frac{1}{2}U+\sum_{i+j=n, ~0\leq i\leq n-1, ~1\leq j\leq n}\lambda^{-i}\tilde{V}^{(-j)},
\label{LP}
\end{eqnarray}
where
\begin{eqnarray}
 \tilde{V}^{(-j)}=\left( \begin{array}{cc} -\frac{1}{2}b_{-j,x} & b_{-j}+\frac{1}{2}-\frac{1}{2\lambda} \\
 m(b_{-j}+\frac{1}{2})\lambda-\frac{1}{2}b_{-j,xx}+\frac{1}{4}(b_{-j}+\frac{1}{2})-\frac{1}{2}m-\frac{1}{8\lambda}&  \frac{1}{2}b_{-j,x} \\ \end{array} \right),
\label{Vj}
\end{eqnarray}
$\lambda$ is the eigenparameter and $b_j$ is defined through equation (4).

By a direct calculation, we obtain the following result.
\begin{proposition}
The n-th CH equation (\ref{CHH}) admits the Lax representation
\begin{eqnarray}
U_{t_{-n}}-V_{x}^{(-n)}+[U,V^{(-n)}]=0,
\label{cc}
\end{eqnarray}
where the Lax pair $U$ and $V^{(-n)}$ given by (\ref{LP}).
\end{proposition}

As $n=1$, we recover the Lax pair of the well-known CH equation (\ref{bCH}) with $\alpha=0$ \cite{CH}
\begin{eqnarray}
\begin{array}{l}
U=\left( \begin{array}{cc} 0 & 1\\
 \frac{1}{4}+\lambda m &  0 \\ \end{array} \right),
 \quad
 V^{(-1)}=\left( \begin{array}{cc}  -\frac{1}{2}b_{-1,x} & b_{-1}-\frac{1}{2\lambda} \\
 mb_{-1}\lambda-\frac{1}{2}b_{-1,xx}+\frac{1}{4}b_{-1}-\frac{1}{2}m-\frac{1}{8\lambda} &  \frac{1}{2}b_{-1,x} \\ \end{array} \right).
 \end{array}
\label{LPCH1}
\end{eqnarray}
As $n=2$, we obtain the Lax pair of the $2$-nd  CH equation (\ref{CHH2})
\begin{eqnarray}
\begin{split}
U=&\left( \begin{array}{cc} 0 & 1\\
 \frac{1}{4}+\lambda m &  0 \\ \end{array} \right),
 \\
 V^{(-2)}=&\left( \begin{array}{cc}  -\frac{1}{2}b_{-2,x} & b_{-2}-\frac{1}{2\lambda} \\
 mb_{-2}\lambda-\frac{1}{2}b_{-2,xx}+\frac{1}{4}b_{-2}-\frac{1}{2}m-\frac{1}{8\lambda} &  \frac{1}{2}b_{-2,x} \\ \end{array} \right)
 \\
 &+\frac{1}{\lambda}\left( \begin{array}{cc}  -\frac{1}{2}b_{-1,x} & b_{-1}+\frac{1}{2}-\frac{1}{2\lambda} \\
 m(b_{-1}+\frac{1}{2})\lambda-\frac{1}{2}b_{-1,xx}+\frac{1}{4}(b_{-1}+\frac{1}{2})-\frac{1}{2}m-\frac{1}{8\lambda} &  \frac{1}{2}b_{-1,x} \\ \end{array} \right).
 \end{split}
\label{LPCH2}
\end{eqnarray}

It has been known that there exist some relations between integrable models in (1+1)-dimensions and  ones in (2+1)-dimensions. For example, assembling of the first two 1+1 dimensional non-trivial members in the AKNS hierarchy: the coupled nonlinear Schr\"{o}dinger equation and the coupled mKdV equation, yields the well-known (2+1)-dimensional KP equation \cite{KSS}-\cite{Cao1}. The compatible solution of the first two members in the KdV hierarchy produces a special solution of the (2+1)-dimensional Sawada-Kotera equation \cite{Cao2}-\cite{SK}.
Here in our paper, we have some similar results listed as follows.
\begin{proposition} Let $t_{-1}=y$, $t_{-2}=t$. Let $m(x,y,t)$ be a compatible solution of the CH equation (\ref{CHH1}) and the 2-nd CH equation (\ref{CHH2}). Then $m(x,y,t)$ provides a special solution to (2+1)-dimensional CH equation (\ref{CH3d}).
In general, if $m(x,t_{-1},t_{-n})$ is a compatible solution of the CH equation (\ref{CHH1}) and the $n$-th CH equation (\ref{CHH}), then the (2+1)-dimensional CH hierarchy (\ref{CHH3d}) has a special solution $m(x,t_{-1},t_{-n})$.
\end{proposition}
The above proposition immediately yields the following corollary.
\begin{corollary} The (2+1)-dimensional CH equation (\ref{CH3d}) possesses a Lax triad $U$, $V^{(-1)}$, $V^{(-2)}$.
In general, the (2+1)-dimensional CH hierarchy (\ref{CHH3d}) possesses a Lax triad $U$, $V^{(-1)}$, $V^{(-n)}$.
\end{corollary}

{\bf Remark 1.} 
Based on proposition 2, we may construct the algebraic-geometric solution of the (2+1)-dimensional CH hierarchy with the method developed in \cite{Cao1,Cao2,Q3}. We will consider this topic in another publication.

\section{The gCH hierarchies in (1+1) and (2+1)-dimensions}

Let us first introduce a pair of Lenard operators \cite{QXL} 
\begin{eqnarray}
J=k_1\partial_{x} m\partial_{x}^{-1}m\partial_{x}+\frac{1}{2}k_2(\partial_x m+m\partial_x), \quad K=\partial_x-\partial^3_x,
\label{gJK}
\end{eqnarray}
and define the Lenard gradients $b_{-k}$ recursively by
\begin{eqnarray}
Kb_{-k}=Jb_{-k+1}, \quad Kb_0=0, \quad k\in\mathbb{Z^{+}}.
\label{gLG}
\end{eqnarray}
We define a gCH hierarchy in (1+1)-dimension as follows
\begin{eqnarray}
\left\{\begin{array}{l}
m_{t_{-n}}=Jb_{-n},\\
Kb_{-j}=Jb_{-j+1},\\
Kb_{-1}=m_{x},
\end{array}\right.  \quad 2\leq j\leq n.
\label{gCHH}
\end{eqnarray}
The first member in (\ref{gCHH}) reads as
\begin{eqnarray}
\left\{\begin{array}{l}
m_{t_{-1}}=\frac{1}{2}k_1\left[m(b_{-1}^2-b_{-1,x}^2)\right]_x+\frac{1}{2}k_2(2m b_{-1,x}+m_xb_{-1}),\\
m=b_{-1}-b_{-1,xx},
\end{array}\right.
\label{gCHH1}
\end{eqnarray}
which is nothing but the gCH equation (\ref{gCH}).
For $n=2$, equation (\ref{gCHH}) is cast into the $2$-nd  gCH equation in the gCH hierarchy (17)
\begin{eqnarray}
\left\{\begin{array}{l}
m_{t_{-2}}=k_1\left[m\partial^{-1}_{x}mb_{-2,x}\right]_x+\frac{1}{2}k_2(2m b_{-2,x}+m_xb_{-2}),\\
b_{-2,x}-b_{-2,xxx}=\frac{1}{2}k_1\left[m(b_{-1}^2-b_{-1,x}^2)\right]_x+\frac{1}{2}k_2(2m b_{-1,x}+m_xb_{-1}),
\\
m=b_{-1}-b_{-1,xx}.
\end{array}\right.
\label{gCHH2}
\end{eqnarray}
For the general case $n\geq 2$, we refer to (\ref{gCHH}) as  the $n$-th gCH equation.

Similar to the (2+1)-dimensional generalization of the CH equation, we extend the (1+1)-dimensional gCH equation (2) to the (2+1)-dimensional system  as follows:
\begin{eqnarray}
\left\{\begin{array}{l}
m_{t}=k_1\left[m\partial^{-1}_{x}mb_{-2,x}\right]_x+\frac{1}{2}k_2(2m b_{-2,x}+m_xb_{-2}),\\
m_y=b_{-2,x}-b_{-2,xxx}.
\end{array}\right.
\label{gCH3d}
\end{eqnarray}
Furthermore, we may define the (2+1)-dimensional gCH hierarchy in the following form:
\begin{eqnarray}
\left\{\begin{array}{l}
m_{t_{-n}}=Jb_{-n},\\
Kb_{-j}=Jb_{-j+1},
\\
m_y=Kb_{-2},
\end{array}\right.  \quad 3\leq j\leq n.
\label{gCHH3d}
\end{eqnarray}
In particular, as $k_1=0$ and $k_2=2$, our (2+1)-dimensional gCH hierarchy (\ref{gCHH3d}) is reduced to the (2+1)-dimensional CH hierarchy (\ref{CHH3d}).

Let us now show that the gCH hierarchies admit Lax representations.
Let 
\begin{eqnarray}
U=\frac{1}{2}\left( \begin{array}{cc} -1 & \lambda m\\
 -k_1\lambda m-k_2\lambda &  1 \\ \end{array} \right),
 \quad
 V^{(-n)}=U+\sum_{0\leq j\leq n-1}\lambda^{-2j}\tilde{V}^{-(n-j)},
\label{gLP}
\end{eqnarray}
where
\begin{eqnarray}
 \tilde{V}^{(-j)}=-\frac{1}{2}\left( \begin{array}{cc} A &
 B \\
 C &
 -A \\ \end{array} \right),
\label{gVj}
\end{eqnarray}
with
\begin{eqnarray}
\begin{split}
A&=\lambda^{-2}+k_1\partial^{-1}mb_{-j,x}+\frac{1}{2}k_2(b_{-j}-b_{-j,x})-1,
\\
B&=-\lambda^{-1}(m-b_{-j,x}+b_{-j,xx})+\lambda m(-k_1\partial^{-1}mb_{-j,x}-\frac{1}{2}k_2b_{-j}+1),
\\
C&=\lambda^{-1}[k_1(m+b_{-j,xx}+b_{-j,x})+k_2]-\lambda (k_1m+k_2)(-k_1\partial^{-1}mb_{-j,x}-\frac{1}{2}k_2b_{-j}+1).
\end{split}
\label{gVjc}
\end{eqnarray}
Direct calculations lead to the following proposition.
\begin{proposition}
The gCH hierarchy (\ref{gCHH}) possesses the Lax representation
\begin{eqnarray*}
U_{t_{-n}}-V_{x}^{(-n)}+[U,V^{(-n)}]=0,
\label{gcc}
\end{eqnarray*}
with the Lax pair $U$ and $V^{(-n)}$ given by (\ref{gLP}).
\end{proposition}

In particular, the Lax pair of the gCH equation (\ref{gCHH1}) are given by
\begin{eqnarray}
\begin{array}{l}
U=\frac{1}{2}\left( \begin{array}{cc} -1 & \lambda m\\
 -k_1\lambda m-k_2\lambda &  1 \\ \end{array} \right),
\quad
 V^{(-1)}=\left( \begin{array}{cc}  A_1 & B_1 \\
 C_1 &  -A_1 \\ \end{array} \right),
 \end{array}
\label{LPgCH1}
\end{eqnarray}
with
\begin{eqnarray}
\begin{split}
A_1&=\lambda^{-2}+\frac{1}{2}k_1(b_{-1}^2-b_{-1,x}^2)+\frac{1}{2}k_2( b_{-1}-b_{-1,x}),
\\
B_1&=-\lambda^{-1}(b_{-1}-b_{-1,x})-\frac{1}{2}\lambda m\left[ k_1(b_{-1}^2-b_{-1,x}^2)+k_2 b_{-1}\right],
\\
C_1&= \lambda^{-1}[k_1(b_{-1}+b_{-1,x})+k_2]+\frac{1}{2}\lambda \left[k_1^2m(b_{-1}^2-b_{-1,x}^2)+k_1k_2(m b_{-1}+b_{-1}^2-b_{-1,x}^2)+k_2^2b_{-1}\right].
\end{split}
\label{gVjc1}
\end{eqnarray}
The Lax pair of the $2$-nd  gCH equation (\ref{gCHH2}) are given by
\begin{eqnarray}
\begin{array}{l}
U=\frac{1}{2}\left( \begin{array}{cc} -1 & \lambda m\\
 -k_1\lambda m-k_2\lambda &  1 \\ \end{array} \right),
\quad
 V^{(-2)}=U+\tilde{V}^{(-2)}+\lambda^{-2}\tilde{V}^{(-1)},
 \end{array}
\label{LPgCH2}
\end{eqnarray}
where $\tilde{V}^{(-1)}$ and $\tilde{V}^{(-2)}$ are defined by (\ref{gVj}) and (\ref{gVjc}).

One may easily check the following results.
\begin{proposition} Let $t_{-1}=y$, $t_{-2}=t$. Let $m(x,y,t)$ be a compatible solution of the gCH equation (\ref{gCHH1}) and the $2$-nd  gCH equation (\ref{gCHH2}). Then $m(x,y,t)$ provides a special solution to (2+1)-dimensional gCH equation (\ref{gCH3d}).
In general, if $m(x,t_{-1},t_{-n})$ is a compatible solution of the gCH equation (\ref{gCHH1}) and the $n$-th gCH equation (\ref{gCHH}), then the (2+1)-dimensional gCH hierarchy (\ref{gCHH3d}) has a special solution $m(x,t_{-1},t_{-n})$.
\end{proposition}

\begin{corollary} The (2+1)-dimensional gCH equation (\ref{gCH3d}) possesses the Lax triad $U$, $V^{(-1)}$, $V^{(-2)}$ given by (\ref{LPgCH1}) and (\ref{LPgCH2}) .
In general, the (2+1)-dimensional gCH hierarchy (\ref{gCHH3d}) possesses the Lax triad $U$, $V^{(-1)}$, $V^{(-n)}$ given by (\ref{gLP}).
\end{corollary}

\section{Peakon solutions to the 2dgCH equation (\ref{gCH3d})}

Assume the single-peakon solution of (2+1)-dimensional gCH equation (\ref{gCH3d}) is given in the form of
\begin{eqnarray}
b_{-2}=p(y,t)e^{-\mid x-q(y,t)\mid}, \quad m=2r(y,t)\delta(x-q(y,t)), \label{ocp}
\end{eqnarray}
where $p(y,t)$, $q(y,t)$ and $r(y,t)$  are to be determined.
Substituting (\ref{ocp}) into (\ref{gCH3d}) and integrating against the test function with support around the peak, we finally arrive at
\begin{eqnarray}
\left\{\begin{array}{l}
r_{y}=r_{t}=0,\\
q_{y}=-\frac{p}{r},\\
q_{t}=-\frac{1}{3}k_1rp-\frac{1}{2}k_2p,
\end{array}\right. \label{socp1}
\end{eqnarray}
which yields
\begin{eqnarray}
\left\{\begin{array}{l}
r=c,\\
q=F(y+(\frac{1}{3}k_1c^2+\frac{1}{2}k_2c)t),\\
p=-cq_y,
\end{array}\right. \label{socp2}
\end{eqnarray}
where $c$ is an arbitrary constant, $F$ is an arbitrary smooth function. Thus, the single-peakon solution of equation (\ref{gCH3d}) is given by
\begin{eqnarray}
\begin{array}{l}
b_{-2}=-cF_y\left(y+(\frac{1}{3}k_1c^2+\frac{1}{2}k_2c)t\right)e^{-\mid x-F\left(y+(\frac{1}{3}k_1c^2+\frac{1}{2}k_2c)t\right)\mid},
\\
m=2c\delta\left(x-F\left(y+(\frac{1}{3}k_1c^2+\frac{1}{2}k_2c)t\right)\right).
\end{array}
\label{ocps}
\end{eqnarray}
As $k_1=0$, $k_2=2$, we recover the single-peakon solution of the $(2+1)$-dimensional CH equation proposed in \cite{EP}.

\begin{figure}
\begin{minipage}[t]{0.5\linewidth}
\centering
\includegraphics[width=2.2in]{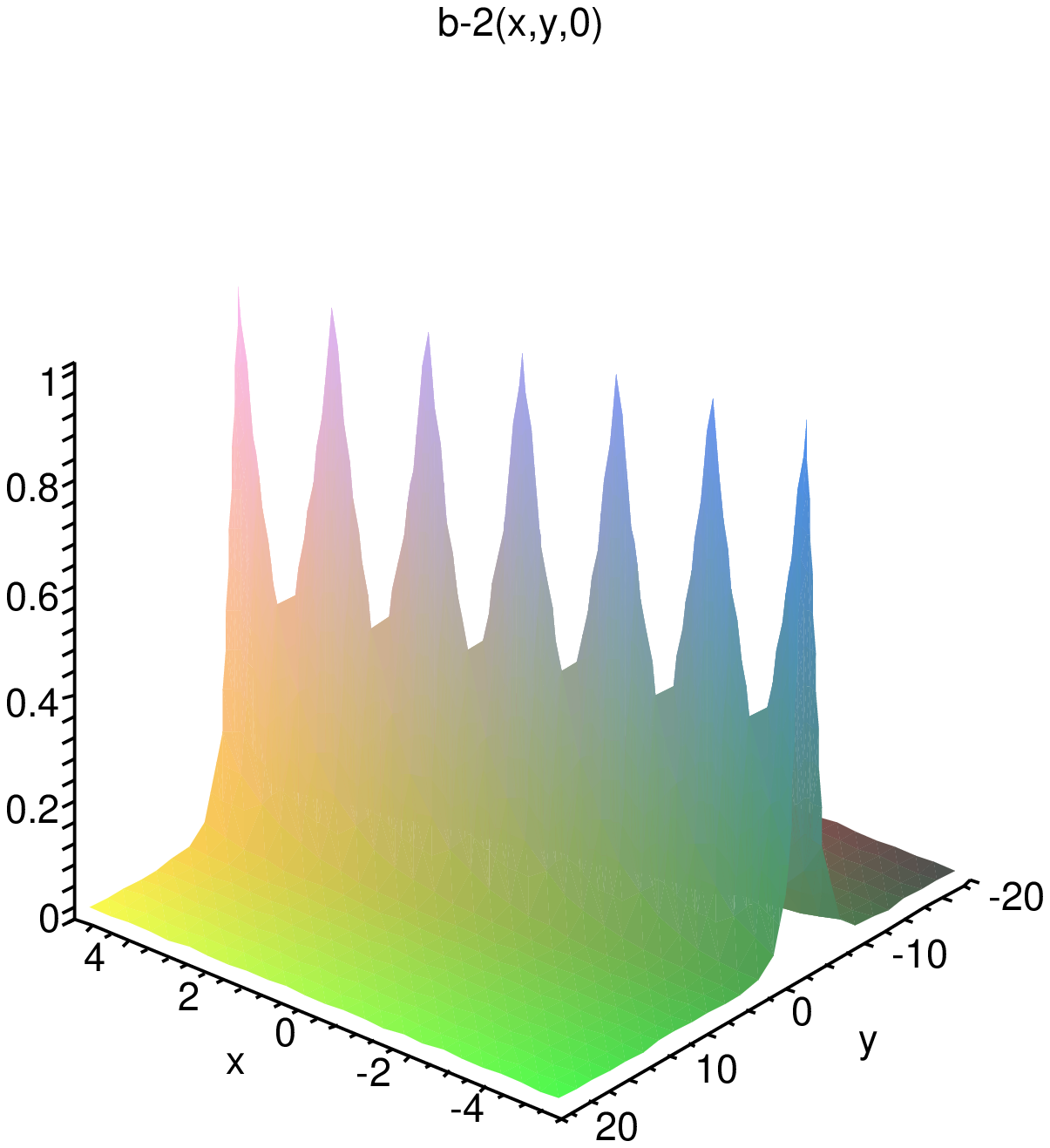}
\caption{\small{Single-peakon solution $b_{-2}(x,y,t)$ in (\ref{ocps2}) with $c=-1$ at $t=0$.}}
\label{f1}
\end{minipage}
\hspace{2.0ex}
\begin{minipage}[t]{0.5\linewidth}
\centering
\includegraphics[width=2.2in]{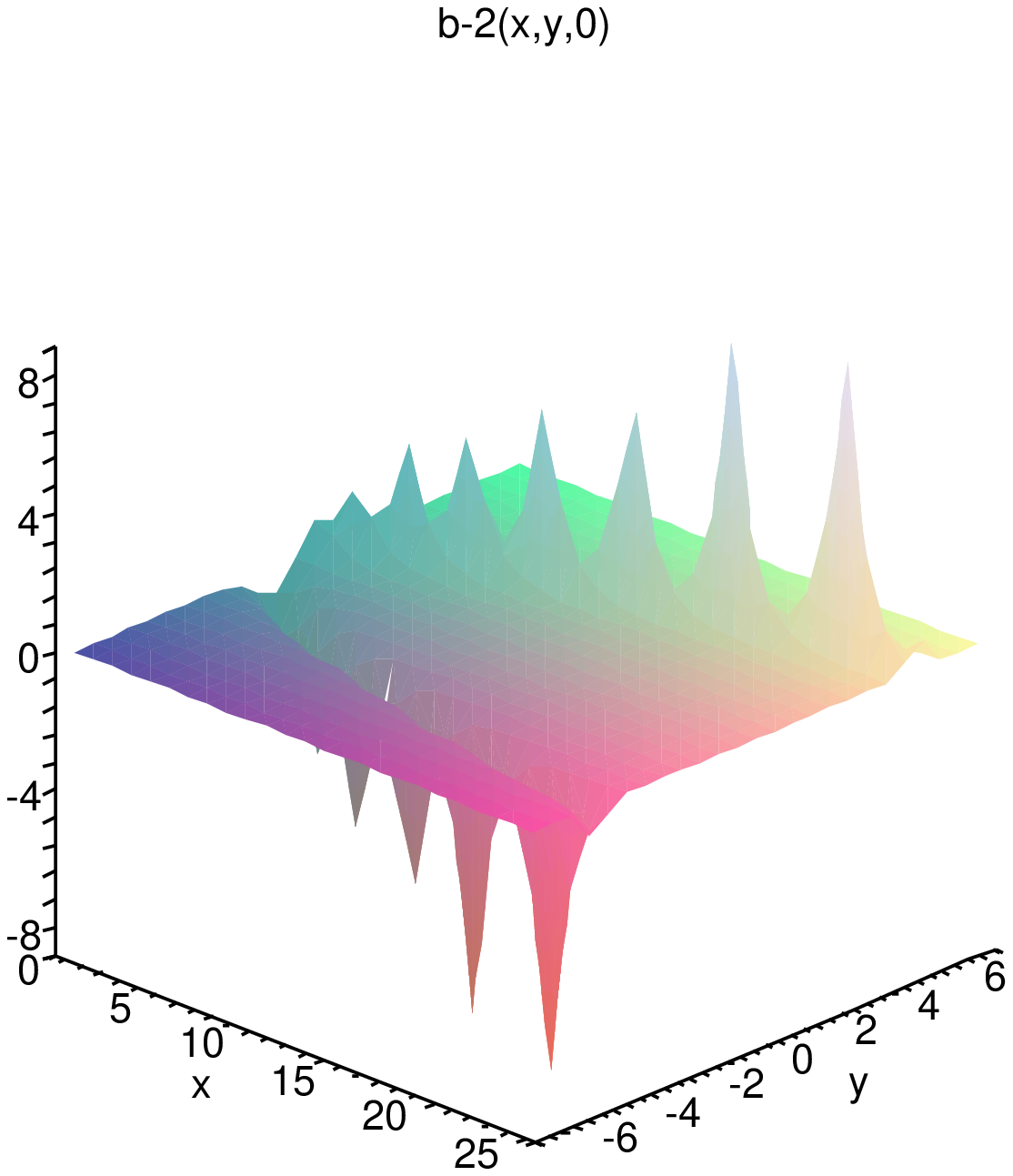}
\caption{\small{Single-peakon solution $b_{-2}(x,y,t)$ in (\ref{ocps3}) with $c=-1$ at $t=0$.}}
\label{f2}
\end{minipage}
\end{figure}

In particular, if we take $F(y+(\frac{1}{3}k_1c^2+\frac{1}{2}k_2c)t)=y+(\frac{1}{3}k_1c^2+\frac{1}{2}k_2c)t$, then the single-peakon solution of equation (\ref{gCH3d}) becomes
\begin{eqnarray}
\begin{array}{l}
b_{-2}=-ce^{-\mid x-y-(\frac{1}{3}k_1c^2+\frac{1}{2}k_2c)t\mid},
\\
m=2c\delta\left(x-y-(\frac{1}{3}k_1c^2+\frac{1}{2}k_2c)t\right).
\end{array}
\label{ocps2}
\end{eqnarray}
See Figure \ref{f1} for the graph of the single-peakon solution $b_{-2}(x,y,t)$ at $t=0$.
If we take $F(y+(\frac{1}{3}k_1c^2+\frac{1}{2}k_2c)t)=\left(y+(\frac{1}{3}k_1c^2+\frac{1}{2}k_2c)t\right)^2$, then the single-peakon solution (\ref{ocps})  becomes
\begin{eqnarray}
\begin{array}{l}
b_{-2}=-2c\left(y+(\frac{1}{3}k_1c^2+\frac{1}{2}k_2c)t\right)e^{-\mid x-\left(y+(\frac{1}{3}k_1c^2+\frac{1}{2}k_2c)t\right)^2\mid},
\\
m=2c\delta\left(x-\left(y+(\frac{1}{3}k_1c^2+\frac{1}{2}k_2c)t\right)^2\right).
\end{array}
\label{ocps3}
\end{eqnarray}
See Figure \ref{f2} for the graph of $b_{-2}(x,y,t)$ in (\ref{ocps3}) at $t=0$.

In general, let us suppose that the $N$-peakon has the following form
\begin{eqnarray}
b_{-2}=\sum_{j=1}^N p_j(y,t)e^{-\mid x-q_j(y,t)\mid}, \quad m=2\sum_{j=1}^N r_j(y,t)\delta\left( x-q_j(y,t)\right).\label{Ncp}
\end{eqnarray}
Similar to the cases of one-peakon but with a lengthy calculation, we are able to obtain the following $N$-peakon dynamical system
\begin{eqnarray}
\begin{split}
r_{j,y}&=0,\\
r_{j,t}&=-\frac{1}{2}k_2r_j\sum_{k=1}^Np_k sgn(q_j-q_k)e^{ -\mid q_j-q_k\mid},\\
p_{j}&=-r_{j}q_{j,y},\\
q_{j,t}&=\frac{1}{6}k_1r_{j}p_{j}-\frac{1}{2}k_2\sum_{k=1}^Np_k e^{ -\mid q_j-q_k\mid}+\frac{1}{2}k_1\sum_{i,k=1}^N  r_ip_k(sgn(q_j-q_i)sgn(q_j-q_k)-1)e^{ -\mid q_j-q_i\mid-\mid q_j-q_k\mid}.
\end{split}   \label{dNcp}
\end{eqnarray}

\section {Conclusions and discussions}
In this letter, we have extended the gCH equation to the hierarchies in (1+1)-dimensions and (2+1)-dimensions. We first show the gCH hierarchies admit Lax representation. Then we show the (2+1)-dimensional gCH equation possesses single peakon solution as well as multi-peakon solutions. Other topics, such as smooth soliton solutions, cuspons, peakon stability, and algebra-geometric solutions, remain to be developed.

\section*{ACKNOWLEDGMENTS}
This work was partially supported by the National Natural Science Foundation of China (Grant Nos. 11301229, 11271168, 11171295, and 61328103), the Natural Science Foundation of the Jiangsu Province (Grant No. BK20130224), the Natural Science Foundation of the Jiangsu Higher Education Institutions of China (Grant No. 13KJB110009), and Qiao also thanks the Haitian Scholar Plan of Dalian University of Technology, the China state administration of foreign experts affairs system under the affiliation of China University of Mining and Technology, and the U.S. Department of Education GAANN project (P200A120256) for their cooperations in conducting the research program.

\vspace{1cm}
\small{

}
\end{document}